\documentclass[aps,prl,twocolumn,groupedaddress,showpacs]{revtex4-1}
\usepackage{graphicx}
 
\def\ni{\noindent}
\def\sig{{\sigma}}

\def\br{\vec{r}}

\def\bg{\vec{g}}

\def\bR{\vec{R}}

\def\bF{\vec{F}}

\def\bff{\vec{f}}

\def\brho{\vec{\rho}}
\def\bphi{\vec{\phi}}

\def\cF{{\cal F}}

\def\cR{{\cal R}}

\usepackage{amsmath}
\DeclareMathOperator{\Tr}{Tr}

\usepackage{color}
\definecolor{endnotes}{RGB}{6,69,173}
\definecolor{crossrefs}{RGB}{128,0,0}
\definecolor{links}{RGB}{6,50,143}

\usepackage[colorlinks=true,linkcolor=,citecolor=endnotes,urlcolor=links]{hyperref}

\begin{document}
%\setlength{\abovedisplayskip}{0pt}
%\setlength{\belowdisplayskip}{0pt}
%\setlength{\abovedisplayshortskip}{0pt}
%\setlength{\belowdisplayshortskip}{2pt}

%@twocolumnfalse\endcsname 
 
\title{Inter-dependence of the volume and stress ensembles and equipartition \\
in statistical mechanics of granular systems}
\author{Raphael Blumenfeld$^{1,2}$, Joe F. Jordan$^1$ and Sam F. Edwards$^2$ }

\affiliation{ $^1$ Earth Science and Engineering and Inst. of Shock Physics, Imperial College, London SW7 2BP, UK\\
$^2$~Biological~and~Systems,~Cavendish~Laboratory,~J.~J.~Thomson~Avenue,~Cambridge~CB3~0HE,~UK } 

\date{\today} 

\begin{abstract} 

\ni We discuss the statistical mechanics of granular matter and derive several significant results. 
First, we show that, contrary to common belief, the volume and stress ensembles are inter-dependent, necessitating the use of both. 
We use the combined ensemble to calculate explicitly expectation values of structural and stress-related quantities for two-dimensional systems. 
We thence demonstrate that structural properties may depend on the angoricity tensor and that stress-based quantities may depend on the compactivity.  
This calls into question previous statistical mechanical analyses of static granular systems and related derivations of expectation values.
Second, we establish the existence of an intriguing equipartition principle - the total volume is shared equally amongst both structural and stress-related degrees of freedom. 
Third, we derive an expression for the compactivity that makes it possible to quantify it from macroscopic measurements.

\end{abstract} 


\maketitle

\ni The statistical mechanical formalism, introduced to describe granular materials\cite{EdOa89,MeEd89,Ed90}, was expected to be a platform for derivations of experimentally measurable equations of state and constitutive relations. It has not yet lived up to its full potential due to several difficulties that traditional thermodynamic theories do not suffer from: lack of ergodicity, uncertainty over the identities and number of degrees of freedom (DoF), and the difficulty to realise a simple analog of a thermometer - a `compactometer'. 
Whilst these problems reflect more on the application of the theory rather than on the theory itself, a more serious concern has arisen from recent suggestions of an absence of an equipartition principle\cite{AlLu02,WaMe08} in agitated systems.  
Here we derive a number of significant results. First, we show that the correct phase space consists of both structural and force DoF, thus  calling into question much of the results in the literature, obtained from either ensemble alone. Second, we show the existence of an equipartition principle in two-dimensional static systems. Third, we show that, in such systems, the compactivity can be quantified directly from macroscopic mean volume measurements.

\ni The initial statistical mechanical approach was based on a volume partition function of $N(>>1)$ grains\cite{EdOa89},

\begin{equation}
Z_v = \int e^{-\frac{W}{X_0} } d\{{\rm all\ structural\ DoF}\} 
\label{Ai}
\end{equation}
where $W$ is a volume function that sums over all the possible volumes that basic volume elements can realise and $X_0$ is the compactivity - a measure of the fluctuations in the ensemble of realisations that is the analogue of temperature. The structural DoF (SDF), identified explicitly below, are all the independent variables that determine the structure of an assembly of grains in mechanical equilibrium, given the mean number of force-carrying contacts per grain $\bar{z}$\cite{Bl08c}. 

\ni The partition function $Z_v$ enables almost closed thermodynamics. For example, once $W$ and the SDF have been identified (see below), the mean volume 2D can be computed. 
Nevertheless, $Z_v$ is unable to specify the macroscopic state of the system completely, since the entropy remains only partially accounted for, it leaves out an entire set of microstates - those of different stress states.
These microstates are described by a different partition function, $Z_f$\cite{EdBl05,BlEd06,Heetal07}, an idea supported later numerically\cite{Heetal07,Puetal10}.
The stress ensemble gives rise to the partition function

\begin{equation}
Z_f = \int e^{-\sum_{\alpha\beta}\frac{1}{X_{\alpha\beta}} \cF_{\alpha\beta}} d\{{\rm all\ boundary\ forces}\}  
\label{Aii}
\end{equation}
Here $\alpha,\beta$ run over the Cartesian components $x,y$ and $\cF_{\alpha\beta}$ are the components of the force moment function, from which the stress $\sig_{\alpha\beta}$ is derived,

\begin{equation}
 \cF_{\alpha\beta} = \sum_{g} V^g \sig^g = \sum_{gg'} \bF_{\alpha}^{gg'}\cR_{\beta}^{gg'}  
\label{Aiii}
\end{equation}
Here the sum runs over pairs of grains in contact $gg'$, $\cR^{gg'}$ is the position of the contact point, measured from the centroid of grain $g$,  $\bF^{gg'}$ is the force that $g'$ applies to $g$ and $V^g$ is the volume associated with grain $g$.
$X_{\alpha\beta}=\partial\cF_{\alpha\beta}/\partial S$ has been named `angoricity'  - a tensorial analogue of the temperature and the compactivity\cite{EdBl05,EdBl07,BlEd09} - and $S$ is the entropy, defined as the log of the number of both structural configurations and stress states. Note that integrating this partition function, any expectations values would be a function of all $\vec{\cR}^{gg'}$.

\ni The volume and stress ensembles are treated in the literature as independent, leading to a total partition function $Z=Z_vZ_f$. Consequently, results have been derived solely from the statistics of one ensemble or the other.   

\ni Here, we challenge this view. We argue that such derivations are misguided and we outline the calculation of the correct partition function. Using the correct ensemble we demonstrate derivation of a number of expectation values in two dimensions (2D), including the expected intergranular force distribution, and derive a surprising equipartition principle for static systems.

\bigskip

\ni We put forward the following three arguments.

\ni 1. {\it The volume ensemble alone is insufficient to describe the entropy of mechanically stable granular systems}. \\
\ni A volume ensemble implies the exact same boundary forces. However, many-grain experiments cannot reproduce the same grain configurations, nor the precise forces on every boundary grain. Only global boundary stresses can be controlled, i.e. averages over boundary force components. Thus, the statistics of the boundary forces must be taken into consideration. 
 
\ni 2. {\it The stress ensemble alone is insufficient to describe the entropy of mechanically stable granular systems}. \\
The stress ensemble comprises a fixed granular configuration, to which all combinations of boundary forces are applied. The ensemble is subject to constraints, e.g. that the total boundary stresses are fixed. Such a system cannot be realised experimentally in very large assemblies (albeit possible in numerical experiments). Indeed, any integration over $Z_f$ remains a function of the SDF.

\ni 3. {\it The volume and stress partition functions are inter-dependent, $Z\neq Z_vZ_f$}. \\
\ni This statement follows from the above two arguments - correct calculations of expectation values must be based on a combined ensemble of all structural arrangements and all boundary forces. Specifically, this is a consequence of the explicit dependence of both the volume function in $Z_v$ and the force moment function in $Z_f$ on the structural DoF (SDF). 

\ni The above arguments hold in any dimension and we proceed to illustrate them explicitly in 2D. Consider an ensemble of 2D $N$-grain systems ($N\gg 1$), each of mean contact number $\bar{z}$. The systems are in mechanical equilibrium under $M$ external compressive forces, acting on the boundary grains. We disregard body forces, in the absence of which `rattlers' can also be ignored, as they do not affect the stress states in static piles.

\begin{figure}["here"]
\includegraphics[width=3.5cm]{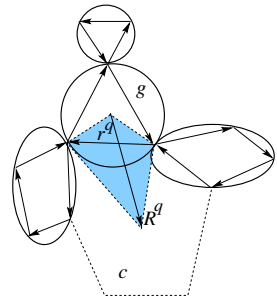}
\caption{The vectors $\br^q$ connect contact points clockwise around grain $g$. The vectors $\bR^q$ connect from grain $g$ centroids to cell $c$ centroid. These vectors are the diagonals of quadron $q$ (blue).}
\label{fig:1}  
\end{figure}

\ni It was proposed to use `quadrons' \cite{BaBl02,BlEd03} as the elementary volumes, both in two and in three dimensions. These are space-tessellating (generically) quadrilateral elements (figure \ref{fig:1}). The quadron is constructed on  two vectors as its diagonals: $\br^q$ connects contact points around a grain in the clockwise direction and $\bR^q$ extends from the centroid (mean position) of the contacts around the grain to the centroid of the contacts around a neighbour cells. 
In terms of these, the volume function is $W=\sum_q v^q= \frac{1}{2} |\br^q\times\bR^q|$ (summation implied over repeated indices) and the partition function is 

\begin{equation}
Z_v = \int e^{-\frac{1}{2X_0} |\br^q\times\bR^q|} \prod_{q=1}^{N_{sdf}/2}dr_x^q dr_y^q 
\label{Aiv}
\end{equation}
Here $N_{sdf}$ is the number of SDF, discussed below.
The vectors $\bR^q$ can be expressed as linear combinations of the $\br^q$ \cite{BlEd06} and, since the latter close loops, only $N\bar{z}/2$ of them are independent\cite{Bl08c,BlEd06}, leading to $N_{sdf}=N\bar{z}$.  
Defining the vector $\brho \equiv \left(r_x^1, r_x^2, ..., r_x^{N\bar{z}/2}, r_y^1, r_y^2, ..., r_y^{N\bar{z}/2}\right)$, $W$ becomes exactly quadratic and we have

\begin{equation}
Z_v =  \int e^{-\frac{1}{2X_0} a^{qp}_{\alpha\beta} r_{\alpha}^q r_{\beta}^p} \prod_{n=1}^{N\bar{z}/2}\prod_{i=1}^2 dr_i^q = 
\int e^{- \frac{1}{2}\brho\cdot A\cdot\brho}  d^{N\bar{z}}\brho 
\label{Avi}
\end{equation}
Here $p,q$ run over quadrons, $\alpha,\beta$ run over vector components $x,y$ and $A$ is a matrix whose elements are 

\[
\left(A\right)^{qp}_{\alpha\beta} = \frac{1}{X_0}\left\{
\begin{array}{l l}
a^{qp}_{xx} & \quad q,p\leq N\bar{z}/2 \\ 
a^{qp}_{xy} & \quad q\leq N\bar{z}/2 \ ,\ p > N\bar{z}/2 \\ 
a^{qp}_{yx} & \quad p\leq N\bar{z}/2 \ ,\ q > N\bar{z}/2 \\ 
a^{qp}_{yy} & \quad q,p > N\bar{z}/2 \\
\end{array} \right.
 \]

\ni We can now evaluate $Z_v$. Assuming a uniform measure of the DoF and that the contribution of very large $\br$ magnitudes is negligible, (\ref{Avi}) can be calculated explicitly

\begin{equation}
Z_v = \sqrt{\frac{\left(2\pi\right)^{N\bar{z}}}{{\rm det}A}}  
\label{Avii}
\end{equation}

\begin{figure}["here"]
\includegraphics[width=3.5cm]{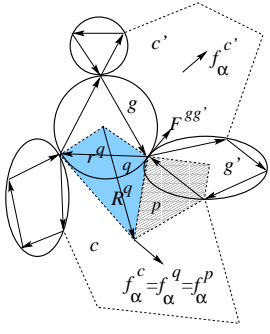}
\caption{$f_{\alpha}^q$ is the $\alpha$ component of the loop force $\bff^c$, which contains the quadron $q$ (shaded blue). The loop forces of $c$ and $c'$ give the inter granular force at the contact point that they share, $\bF^{gg'}=\bff^c - \bff^{c'}$\cite{BaBl02}. Quadron $p$ (striped) shares the same loop as $q$ and hence also the same loop force.}
\label{fig:3}  
\end{figure}

\ni The stress partition function consists of all the possible combinations of compressive forces on the boundary grains, $\bg^m$ ($m=1,2,...,M$), subject to the constraint that the total stress on the boundary is  fixed \cite{EdBl07,Bl08c,BlEd09}. Since the configuration is presumed fixed, only boundary forces that do not drive the system out of mechanical equilibrium are permissible, i.e. the boundary stresses must be below the yield surface. 
It has been shown\cite{BaBl02,BlEd03} that the force moment function (\ref{Aiii}) can be written as 

\begin{equation}
 \cF_{\alpha\beta} = \sum_{gc} f_{\alpha}^q r_{\beta}^q 
\label{Aviii}
\end{equation}
where $f_{\alpha}^q$ is the $\alpha$ component of a loop force of the cell containing the quadron $q$ (see figure \ref{fig:3}). The loop forces are defined in terms of the contact forces\cite{BaBl02}, e.g. $\bF^{gg'}=\bff^c - \bff^{c'}$ in figure \ref{fig:3}, and they conveniently satisfy force balance conditions. In 2D assemblies of arbitrarily-shaped frictional grains, there are $N\left(\bar{z}/2-1\right)$ loop forces (a consequence of Euler relation), of which $N/2$ can be determined from the torque balance conditions. 
For clarity, we next focus on isostatic systems ($\bar{z}=3$); extension to hyperstatic assemblies ($\bar{z}>3$) is possible. 
Substituting (\ref{Aviii}) into (\ref{Aii}) gives

\begin{equation}
Z_f = \int e^{-\frac{1}{X_{\alpha\beta}} f^q_{\alpha} r^q_{\beta}} \prod_{m=1}^{M} d\bg^m  
\label{Aix}
\end{equation}
where the integration runs over all the independent boundary forces  $\bg^m$. 
Since quadrons sharing the same cell have the same loop force (figure \ref{fig:3}) and the loop forces depend linearly on the $M$ boundary forces, then only $N/2$ of the $N\bar{z}$ quadron forces  are independent. It is therefore convenient to define the loop forces vector $\bphi \equiv \left(f_x^1, f_x^2, ..., f_x^{N/2}, f_y^1, f_x^2, ..., f_y^{N/2}\right)$. The solution for $\bphi$ in terms of the boundary forces can be written as

\begin{equation}
\phi^c_{\alpha} = C^{qm}_{\alpha\beta}g^m_{\beta} 
\label{Ax}
\end{equation} 
where $\alpha,\beta=x,y$, $c=1,2,\dots,N/2$ runs over all cells, $m = 1,2,\dots,M$ runs over all boundary forces and $C$ is $N\times 2M$ matrix. In terms of these, $\bff^q=E\bphi$, where $E$ is a $N\bar{z}\times N$ matrix.
Defining further  $B^{qp}_{\alpha\beta}=X^{-1}_{\alpha\beta}\delta_{qp}$, with $\delta_{qp}$ being the delta function, we finally obtain $Z_f$ 

\begin{equation}
Z_f = \int e^{-\bphi\cdot E^T \cdot B\cdot\brho} \prod_{m=1}^{M} d^2\bg^m  = 
\int e^{-\bg\cdot C^T\cdot E^T \cdot B\cdot\brho} \prod_{m=1}^{M} d^2\bg^m  
\label{Axi}
\end{equation}

\ni To compute the total partition function, we need to integrate over the combined volume-stress phase space, $dZ = dZ_v dZ_f$

\begin{eqnarray}
Z & = & \int e^{- \frac{1}{2} \brho\cdot A\cdot\brho - \bg\cdot Q^T \cdot\brho} \left( d^{N\bar{z}}\brho\right) \left( d^{2M} \bg \right)
\label{Axiii}
\end{eqnarray}
where we have defined, for brevity, $Q=\! B^T\! \cdot\!  E\! \cdot\!  C$.
This expression establishes our claim that $Z$ is {\it not} the product of $Z_v$ and $Z_f$ of eqs. (\ref{Avii}) and (\ref{Axi}).  

\ni Integrating (\ref{Axiii}) is straightforward due to the integrand's Gaussian form and we use it next to calculate several expectation values. 
The exponential contains a linear and a quadratic term in $\brho$ and, completing to quadrature and changing variables to $\tilde{\rho} = \rho + A^{-1} Q \bg$, we can separate the variables in the exponent, 

\begin{equation}
 - \frac{1}{2} \brho \, A \, \brho \,- \bg \, Q^T\, \brho \, = \, - \frac{1}{2} \vec{\tilde{\rho}} \, A \, \vec{\tilde{\rho}}  \,+\, \frac{1}{2} \bg\, P\, \bg
\label{conjugate_variables}
\end{equation}
where we define for shorter notation $P=\! Q^T\! \cdot\! A^{-1}\! \cdot\!  Q$.
Calculating the mean volume we obtain:

\begin{equation*}
\langle V \rangle =  \frac{X_0}{2Z} \cdot \int \left( \vec{\tilde{\rho}}  \, A \, \vec{\tilde{\rho}}  \,+\, \bg\, P\, \bg \right) 
  e^{\frac{1}{2}\left(-\vec{\tilde{\rho}}  \, A \, \vec{\tilde{\rho}}  \,+\, \bg\, P\, \bg \right)} 
  \left( d^{N\bar{z}}\vec{\tilde{\rho}} \right) \left( d^{2M} \bg \right)
\end{equation*}
which separates into two gaussian integrals, giving

\begin{equation}
  \langle V \rangle = 
  \frac{ \bar{z}N + 2M }{2} X_0 
  \label{expvalvolume}
\end{equation}

\ni This result is significant for several reasons. First, it is independent of the details of the connectivity matrix $A$ and of the particular stress state.  
Second, it reveals a striking equipartition principle: the mean volume is shared equally among the $\bar{z}N$ structural and the $2M$ force DoF, each having on average $X_0/2$. It is analogous to the mean energy per DoF in thermal systems $3k_BT/2$, but we emphasize that no energy is involved in this formalism.
Third,  it quantifies the compactivity $X_0$ in terms of measurable quantities. 
An important consequence of this finding is that it makes possible to start analyses `inductively' by \emph{assuming} that the volume per DoF is $X_0/2$, as done as standard in thermal systems analogously with $k_B T$. 
Note that using only $Z_v$ (eq.~\ref{Avi}) gives $\langle V_v \rangle = \bar{z} N X_0/2$, which overestimates the compactivity.

\ni All relevant expectation values can be expressed in terms of $ \vec{\tilde{\rho}} $ and $\bg$ and hence evaluated, albeit with more algebra,

\begin{eqnarray}
\langle \cF_{\alpha\beta} \rangle & = & - \frac{\partial \ln{Z}}{\partial \left(1/X_{\alpha\beta}\right)} = \sum_i^{2M} \frac{R^{\alpha\beta}_{ii}}{p_i}  \label{expvalforcemom} \\
\langle \brho\cdot\brho \rangle & = & - \frac{\partial \ln{Z}}{\partial A_{ii}} =  \Tr{A^{-1}} + \sum_i^{2M} \frac{T_{ii}}{p_i}  \label{expvalrhorho} \\
\langle \bff\cdot\bff \rangle & = & - \eta_{ij}\frac{\partial \ln{Z}}{\partial P_{ij}} = \sum_i^{2M} \frac{U_{ii}}{p_i}  \label{expvalforceforce}
\end{eqnarray}
where $R=Y^T\cdot C^T\cdot E^T\otimes A^{-1}\cdot Q\cdot Y$, $T=Y^T\cdot Q^T\cdot A^{-1}\cdot A^{-1}\cdot Q\cdot Y$, 
$U=Y^T\cdot C^T\cdot E^T\cdot E\cdot C\cdot Y$, $Y$ is the matrix that diagonalises $P$, $p_i$ are the eigenvalues of $P$, and $\eta_{ij}$ are straightforward functions of $E$ and $C$. 
Note that both results (\ref{expvalvolume}) and (\ref{expvalforceforce}) are directly relevant to experimental measurements\cite {MaBe05,Loetal09}.
\ni These exact results do more than demonstrate the utility of the combined ensemble, they reveal unexpected dependences of these expectation values on the compactivity and angoricity. 
For example,  $\langle \brho\cdot\brho \rangle$ is not only proportional to $X_0$, as one would expect, but it also depends on $X_{\alpha\beta}$ via a homogeneous function (HF) of order $0$. 
Also unexpectedly, the mean inter-granular force magnitude square $\langle \bff\cdot\bff \rangle$ is both a HF of order $2$ of $X_{\alpha\beta}$ and linear in $X_0$. 
Yet $\langle \cF_{\alpha\beta} \rangle$ is, unsurprisingly, a HF of order $1$ of the angoricity and independent of $X_0$. 
These results show the significance of using  both the stress and the volume ensembles.

\ni To conclude, we have presented three main results.
First, we have shown that the compactivity-based volume ensemble and the angoricity-based stress ensemble are dependent and need to be used simultaneously. We reiterate, the entropy is the log of all the microstates, which include both the SDF and the stress states -- because $Z_v$ and $Z_f$ are dependent, it is not simply the sum of the configurational and stress entropies. This calls into question the large body of work, obtained from either ensemble alone. 
We have used the combined partition function to obtain explicitly the expectation values of: the mean volume, force moment, distance between intra-grain neighbour contact points, and contact force magnitude. We find, surprisingly, that $\langle V\rangle$ depends explicitly on the force degrees of freedom, that $\langle \rho\cdot\rho\rangle$ depends on the angoricity, and that $\langle \bff \cdot \bff \rangle$ on the compactivity. 
Second, the calculation of $\langle V\rangle$ reveals the existence of an equipartition principle -  the mean volume of static systems is shared equally amongst both structural and force-related DoF, with each getting a volume of $X_0/2$.
This result shows that, although equipartition is questionable in dynamic dense systems\cite{AlLu02,WaMe08}, it exists for static ones. 
Moreover, since static granular systems are the equivalent of ``zero temperature'' granular fluids, this result gives hope that an equipartition principle may be found for dense dynamic systems by extending dynamic descriptions to include structural and force DoF.
Third, we have derived an expression for the compactivity in terms of measurable quantities - the mean volume and the mean contact number and the loading forces.
A significant implication of our results is that the compactivity and angoricity are not in themselves the conjugate variables of volume and force moment, as previously believed. Instead, it is the expression in eq.~(\ref{conjugate_variables}) that represents a convolution of the volume and force moment functions with the compactivity and angoricity.

\ni It would be interesting to test our analysis experimentally and numerically, e.g. by constructing assemblies at different compactivities and angoricities and examining expectation values as functions of these parameters.  
Furthermore, since the arguments establishing the inter-dependence of the volume and stress ensembles hold in any dimension, it should be possible to extend our analysis to 3D systems, at least numerically. 
 
\smallskip

\subsection{\bf Acknowledgement}

\ni This work has been funded by EPSRC - EP/H051716/1

%\href{http://dx.doi.org/}{CTAN}

\end{document}